\documentclass[12pt]{iopart}

\usepackage{hyperref}
\usepackage{iopams}  
\usepackage[utf8]{inputenc} 			
\usepackage[T1]{fontenc}				
\usepackage{graphicx}
\usepackage{float}
\usepackage[usenames,dvipsnames]{xcolor}
\usepackage{tikz}
\usepackage{ulem}
\renewcommand{\vec}[1]{\mathbf{#1}}
\newcommand{\Gvec}[1]{\boldsymbol{#1}}

\newcommand*\colvec[3][]{
    \begin{pmatrix}\ifx\relax#1\relax\else#1\\\fi#2\\#3\end{pmatrix}
}

\begin{document}

\title{On the Einstein relation between mobility and diffusion coefficient in an active bath}

\author{Alexandre Solon}
\address{Sorbonne Universit\'e, CNRS, Laboratoire de Physique Th\'eorique de la Mati\`ere Condens\'ee, LPTMC, F-75005 Paris, France}
\ead{alexandre.solon@sorbonne-universite.fr}

\author{Jordan M. Horowitz}
\address{
Department of Biophysics, University of Michigan, Ann Arbor, Michigan, 48109, USA\\
Center for the Study of Complex Systems, University of Michigan, Ann Arbor, Michigan 48104, USA\\
Department of Physics, University of Michigan, Ann Arbor, Michigan, 48109, USA}
\ead{jmhorow@umich.edu}


\begin{abstract}
  An active bath, made of self-propelling units, is a nonequilibrium
  medium in which the Einstein relation $D=\mu k_BT$ between the
  mobility $\mu$ and the diffusivity $D$ of a tracer particle cannot
  be expected to hold {\it a priori}. We consider here heavy tracers
  for which these coefficients can be related to correlation functions
  which we estimate. We show that, to a good approximation, an
  Einstein relation does hold in an active bath upon using a different
  temperature which is defined mechanically, through the pressure
  exerted on the tracer.
\end{abstract}

Since the seminal experiments of Wu and
Libchaber~\cite{wu_particle_2000}, the diffusion of a tracer particle
is known to be enhanced when it is placed in an active bath composed
of self-propelled entities (bacteria in the experiment). Quantifying
this effect is important to understand biological processes such as
transport within a cell and to take advantage of the enhanced mixing
at the microscopic scale due to the active
bath~\cite{kim_controlled_2007,belan_pair_2019}. It has given rise to
many
experimental~\cite{chen_fluctuations_2007,leptos_dynamics_2009,valeriani_colloids_2011,mino_enhanced_2011,jeanneret_entrainment_2016,maggi_memory-less_2017,ortlieb_statistics_2019,lagarde_colloidal_2020,seyforth2021non}
and theoretical
works~\cite{squires_simple_2005,underhill_diffusion_2008,lin_stirring_2011,pushkin_fluid_2013,morozov_enhanced_2014,thiffeault_distribution_2015,reichhardt2015active,burkholder_tracer_2017,dal2019linear,knevzevic2021oscillatory}. In
the case of bacterial baths, an enhanced diffusivity due to either
direct
collisions~\cite{squires_simple_2005,burkholder_tracer_2017,lagarde_colloidal_2020}
or far-field hydrodynamic
interactions~\cite{underhill_diffusion_2008,lin_stirring_2011,pushkin_fluid_2013,morozov_enhanced_2014,thiffeault_distribution_2015}
has been proposed to account for the experimental measurements.


Whatever the type of interactions, the diffusivity of the tracer is
strongly enhanced --- {\it e.g.} by two orders of magnitude in
Ref.~\cite{wu_particle_2000} --- while its mobility is not affected
much. Indeed, most active suspensions are relatively dilute so that
the drag force exerted by the active particles is small compared to
that exerted by the surrounding fluid. Diffusion and mobility thus
have different origins, respectively in the active particles and the
surrounding fluid, so that we do not expect them to be related by an
Einstein relation as in equilibrium.  However, we do expect that if we
look at the drag force and noise imparted only by the surrounding
fluid, they will be related via the fluctuation-dissipation theorem.
Similarly, the active bath imparts drag and noise on the tracer,
though their relationship is not as simple since the active bath is
out of equilibrium.  Multiple methods have been developed to
understand the effect of the active bath, for example using modern
developments in nonequilibrium linear response theory for weakly
interacting tracers~\cite{Maes2014,Steffenoni2016,Mae2020} or by a
perturbative analysis of the stochastic equations of motion for soft
tracers~\cite{Demery2014,Demery2019,Feng2021}.  However, in the the
experimentally relevant limit of hard, strongly interacting tracers,
the effect of an active bath on the diffusion and mobility remains
largely unexplored.



With this in mind, we study in this article the effect of an active
bath on both diffusion and mobility. To this end, we consider an
underdamped tracer subject to passive noise and damping from a
surrounding fluid and to the collisions with active Brownian
particles.  We work in the limit of heavy tracers, so that standard
projection operator methods~\cite{Gardiner,Zwanzig} allow us to write
the the noise and damping due to the active bath as fixed-tracer
correlation functions, which we evaluate. To highlight similarities
and differences, we first consider a bath of passive Brownian
particles for which the Einstein relation $D=\mu k_B T$ between the
mobility $\mu$ and diffusion coefficient $D$ of the tracer directly
follows from the Boltzmann distribution. However, an alternate
derivation based only on mechanical quantities is possible.  From this
alternative perspective, the origin of the temperature in the Einstein
relation comes from the ideal gas law, where it plays the role of the
proportionality constant between pressure and density:
$\Pi=\rho_0 k_B T$.  In the active case, along similar lines we can
introduce a mechanically-defined ``active temperature'' $T_a$ via the
relationship between pressure and density, $T_a=\Pi/(\rho_0 k_B)$,
where $\Pi$ is now the mechanical pressure exerted by the active bath
on the tracer.  We find that, to a good approximation that becomes
exact for large tracers, the damping and noise due to the active bath
obey an Einstein-like relation involving the active temperature. The
full $D$ and $\mu$ are then related by a combination of active and
passive contributions.


In this study, we consider only spherical tracers. For tracers with a
different shape, the physics is expected to be different since they
generically induce long-range (power-law decaying) disturbances in an
active bath~\cite{baek_generic_2018,granek_bodies_2020}. If the tracer
has a polar shape, it will even be spontaneously propelled by the
bath, an effect well-established in
experiments~\cite{di_leonardo_bacterial_2010,sokolov_swimming_2010,kaiser_transport_2014}. We
work here in two or three spatial dimensions (all simulations are in
$d=2$). Specific effects come into play in $d=1$ that have been
recently explored in
Ref.~\cite{banerjee_tracer_2021,granek_anomalous_2021}.

The paper is organized as follows: In Sec.~\ref{sec:model} we
introduce the microscopic model. In Sec.~\ref{sec:eff-dynamics} we
relate the damping and noise due to the bath to correlation functions
involving the force on a fixed tracer which we compute in
Sec.~\ref{sec:FF}. Finally in Sec.~\ref{sec:D-mu} we compute the
diffusivity and mobility of the tracer and conclude with a discussion
in Sec.~\ref{sec:discussion}.

\section{The model}
\label{sec:model}
We consider a tracer with position and velocity ($\vec R, \vec V$) in a
bath of either passive Brownian particles (PBPs) or active Brownian
particles (ABPs). The ensemble is itself in a surrounding fluid at
temperature $T$ that induces a friction of coefficient $\gamma_{T}$
and $\gamma_{B}$ on the tracer and bath particles respectively.

The tracer moves according to the Langevin equation
\begin{equation}
  \label{eq:tracer-dynamics}
 m \dot\vec V=\vec F-\gamma_{T} \vec V+\sqrt{2D_T}\Gvec\xi(t)
\end{equation}
with Gaussian white noise
$\langle\xi_\alpha(t)\xi_\beta(0)\rangle=\delta(t)\delta_{\alpha\beta}$
and $D_T=T \gamma_T$ the noise strength due to the surrounding fluid
(here and henceforth we use units such that $k_B=1$).
$\vec F=\sum_i \vec F_i$ is the total force imparted by the bath
particles, which is the sum of all the forces $\vec F_i$ due to each
particle $i$.

We assume the bath particles to have an overdamped dynamics. Each bath
particle $i$ then follows either one of the two dynamics
\begin{eqnarray}\label{eq:bath-passive}
  \dot\vec r_i&=-\mu_B \vec F_i+\mu_B\vec f_i+\sqrt{2 D_B}\Gvec\xi_i(t), \qquad{\rm (PBP)} \\
  \label{eq:bath-active}
  \dot\vec r_i&=-\mu_B \vec F_i+\mu_B\vec f_i+v_0 \vec u_i , \qquad{\rm (ABP)}
\end{eqnarray}
where $\vec f_i$ is the force exerted by the other bath particles on
particle $i$. In the passive case, the particle feels a Gaussian white
noise with
$\langle\xi_{i\alpha}(t)\xi_{j\beta}(0)\rangle=\delta(t)\delta_{\alpha\beta}\delta_{ij}$
and $D_B=\mu_B k_B T$ with $\mu_B=1/\gamma_B$ while in the active case
it is replaced by a self-propulsion at speed $v_0$ in direction
$\vec u_i$ performing rotational diffusion on the unit sphere. In
$d=2$, this simply reads $\vec u_i=(\cos\theta_i,\sin\theta_i)$ and
$\dot\theta_i=\sqrt{2 D_r}\eta_i(t)$ with $\eta_i$ a delta-correlated
unit-variance Gaussian white noise and $D_r$ the rotational diffusion
coefficient. Note that in general the rotational diffusion can stem
from the active dynamics and thus $D_r$ is not related to the
temperature.

We consider hardcore interactions between the tracer and bath
particles, implemented using the algorithm of
Ref.~\cite{scala_event-driven_2007}. On the contrary, among
themselves, bath particles interact via a truncated harmonic
potential, $\vec f_i=-\nabla_{r_i}V$ with
$V=\sum_{i< j}\frac{k}{2}(\sigma_B-|\vec r_i-\vec r_j|)^2$ if
$|\vec r_i-\vec r_j|<\sigma_B$ and $V=0$ otherwise. This allows us to
vary the bath transport properties by tuning the interaction strength
$k$ while we keep the number density $\rho_0=1$ fixed. Without loss of
generality, we choose the interaction radius of a bath particle
$\sigma_B=1$, thereby fixing the length unit and work in energy units
such that $k_B=1$. The interaction radius between the tracer and the
bath particles is denoted $\sigma$ and is a parameter of the model.
We choose the time unit such that $D_B=1$ for both PBPs and ABPs. In
the later case, $D_B=v^2/(2D_r)=v l_p /2$ where $l_p=v/D_r$ is the
persistence length. We fix $v=2$ and $D_r=0.5$ such that both $D_B$
and $l_p$ are unity.

\section{Effective tracer dynamics}
\label{sec:eff-dynamics}
When the motion of the tracer is slow compared to the bath relaxation time, the effect of the bath on the tracer dynamics can be included in an effective equation of motion.
To derive this reduced equation for the tracer, we use standard projection operator techniques~\cite{Gardiner, Zwanzig,Espanol2002}.
Details can be found in \ref{sec:effective-bath}.
Here, we outline the main steps, following closely the original exposition for equilibrium fluids \cite{vankampen1986}.
 
 We begin with the Fokker-Planck equation for the probability distribution of the joint position of the tracer  in its phase space $({\mathbf R},{\mathbf P})$ and the positions of the  overdamped Brownian bath particles $\{{\bf r}_i\}$~\cite{Gardiner},
\begin{equation}\label{eq:fp-main}
\partial_t P({\mathbf R},{\mathbf P},{\mathbf r}_i) ={\mathcal L}_T P({\mathbf R},{\mathbf P},{\mathbf r}_i) + {\mathcal L}_B  P({\mathbf R},{\mathbf P},{\mathbf r}_i), 
\end{equation}
where the Fokker-Planck operators ${\mathcal L}_T$ and ${\mathcal L}_B$ generate the tracer dynamics and the bath dynamics, respectively, in accordance with the Langevin equations in Eqs.~\ref{eq:tracer-dynamics}, \ref{eq:bath-passive} and \ref{eq:bath-active}.

To extract the slow tracer dynamics, we introduce an operator ${\mathcal P}$ that removes the bath (${\mathcal L}_B{\mathcal P}={\mathcal P}{\mathcal L}_B=0$) by projecting the bath onto its steady-state distribution conditioned on the position of the slow tracer $\pi_B({\mathbf r}_i|{\mathbf R})$, which is defined as the solution of ${\mathcal L}_B\pi_B({\mathbf r}_i|{\mathbf R})=0$:
\begin{equation}
{\mathcal P}P({\mathbf R},{\mathbf P},{\mathbf r}_i)=\pi_B({\mathbf r}_i|{\mathbf R})\int \prod_id{\mathbf r}_iP({\mathbf R},{\mathbf P},{\mathbf r}_i).
\end{equation}
By applying ${\mathcal P}$ and the orthogonal projector ${\mathcal Q}={\mathcal I}-{\mathcal P}$ onto the Fokker-Planck equation \ref{eq:fp-main}, we arrive at a pair of coupled equations for the relevant ${\mathcal P}P$ and irrelevant ${\mathcal Q}P$ parts of the distribution
\begin{eqnarray}\label{eq:Pp}
&\partial_t{\mathcal P}P={\mathcal P}{\mathcal L}_T{\mathcal P}P+{\mathcal P}{\mathcal L}_T{\mathcal Q}P\\
&\partial_t{\mathcal Q}P={\mathcal Q}{\mathcal L}_B{\mathcal Q}P+{\mathcal Q}{\mathcal L}_T{\mathcal P}P+{\mathcal Q}{\mathcal L}_T{\mathcal Q}P.
\end{eqnarray} 
Systematically solving for the irrelevant part ${\mathcal Q}P$ assuming that the bath relaxation is fast results in a closed equation for the tracer's evolution,
\begin{equation}
\partial_t{\mathcal P}P={\mathcal P}{\mathcal L}_T{\mathcal P}P+{\mathcal P}{\mathcal L}_T{\mathcal Q}\int_0^\infty ds\ e^{{\mathcal L}_Bs}{\mathcal Q}{\mathcal L}_T{\mathcal P}P.
\end{equation}
The remainder of the derivation requires evaluating each sequence of operators, assuming spherically symmetric particles.

In the end, we find that the effect of the bath can be encapsulated by an additional Gaussian white noise with strength $D_p$ and friction $\gamma_p$ (the subscript ``p'' stands for
``projected'') such that
\begin{equation}
  \label{eq:tracer-dynamics-effective}
 m \dot \vec V=-(\gamma_T+\gamma_p) \vec V+\sqrt{2(D_T+D_p)}\Gvec\xi(t)
\end{equation}
with coefficients given by the integrals of two-time correlation
functions
\begin{equation}
  \label{eq:coeff_proj-operator}
 \gamma_p\equiv-\frac{1}{d}\int_0^\infty\langle \vec F(t)\cdot\nabla\log \pi_B(0)\rangle_B dt; \qquad D_p\equiv \frac{1}{d}\int_0^\infty\langle \vec F(t)\cdot\vec F(0)\rangle_B dt,
\end{equation}
where $\langle \cdot\rangle_B$ is the average over the bath's
steady-state distribution $\pi_B$ with tracer {\it fixed} at the
origin, and $d$ the dimension of space.

To characterize the tracer dynamics we investigate two
experimentally-accessible parameters, the mobility $\mu$ and the
diffusivity $D$.  The mobility is defined as the response to a small
constant force (say along the $x$-axis) $\vec f=f\vec e_x$,
\begin{equation}
  \label{eq:mu-def}
  \mu\equiv \lim_{f\to 0}\frac{\langle V_x\rangle_f}{f},
\end{equation}
where $\langle\cdot\rangle_f$ is the steady-state average in presence
of the pulling force.  The diffusion coefficient is defined as the
rate of growth of the mean squared displacement
\begin{equation}
  \label{eq:D-deff}
  D\equiv\lim_{t\to\infty} \frac{1}{2 d t} \langle (\vec R(t)-\vec R(0))^2\rangle=\frac{1}{d}\int_0^\infty\langle  \vec V(t)\cdot \vec V(0)\rangle dt.
\end{equation}

For the effective dynamics in
Eq.~(\ref{eq:tracer-dynamics-effective}), we can obtain analytic
expressions for these quantities. It is first straightforward to show
that $\mu=1/(\gamma_T+\gamma_p)$. Furthermore, the steady-state
solution of Eq.~(\ref{eq:tracer-dynamics-effective}) is like a Maxwell
distribution for the tracer velocity, except with an effective
temperature determined by kinematic parameters:
\begin{equation}
  \label{eq:betaeff}
  \pi_V(\vec V)\propto e^{-\frac{m V^2}{2 T_{\rm eff}}};\qquad T_{\rm eff}\equiv \frac{D_T+D_p}{\gamma_T+\gamma_p}.
\end{equation}
An Einstein-like relation then follows from a first-order perturbation
theory. Indeed, linear response directly gives
that~\cite{risken1996fokker}
\begin{equation}
  \label{eq:mu-pert}
  \mu=-\frac{1}{m}\int_0^\infty\left\langle V_x(t)\frac{\partial \pi_V(0)}{\partial V_x}\right\rangle dt.
\end{equation}
One then obtains, using the distribution of Eq.~(\ref{eq:betaeff}),
\begin{equation}
  \label{eq:mu-def2}
  \mu=\beta_{\rm eff}\int_0^\infty\langle V_x(t)V_ x(0)\rangle dt=D/T_{\rm eff},
\end{equation}
where the last equality follows from the spherical symmetry of the
steady state.

All in all, we see that the knowledge of the bath coefficients
$\gamma_p$ and $D_p$ is enough to compute $\mu$, $T_{\rm eff}$ and
thus $D$ from Eq.~(\ref{eq:mu-def2}). To this end, we study in
Sec.~\ref{sec:FF} the correlators that appear in the definitions of
$\gamma_p$ and $D_p$ before turning to $D$ and $\mu$ in
Sec.~\ref{sec:D-mu}.

\section{Force autocorrelation}
\label{sec:FF}

The friction $\gamma_p$ and noise strength $D_p$ characterizing the bath,
defined in Eq.~(\ref{eq:coeff_proj-operator}), are expressed as the
time integral of the correlation functions
$\langle \vec F(t)\cdot\nabla\log \pi_B(0)\rangle_B$ and
$\langle \vec F(t)\cdot\vec F(0)\rangle_B$ respectively. 

For a passive bath, one can use the Boltzmann distribution to relate
the two correlation functions. Let us denote by $V_{\rm HC}$ the
hardcore potential between the tracer and the bath particles. Then
using that at equilibrium $\pi_B\propto e^{-V_{\rm HC}/T}$, we find
that $\nabla\log \pi_B=-\frac{1}{T}\nabla V_{\rm HC}=\vec F$ so that
the two correlation functions are proportional with a factor $T$.
After integration, one recovers the fluctuation-dissipation theorem
$D_p=T\gamma_p$.  The implication of this calculation is that to
determine the two coefficients, $\gamma_p$ and $D_p$, it is enough to
compute the force autocorrelation, which we do in
Sec.~\ref{sec:FF-passive}.

For active baths, which we discuss in Sec.~\ref{sec:FF-active}, the
above reasoning does not hold since the bath particles are not
distributed according to the Boltzmann distribution. Nonetheless, for
non-interacting active particles, we find that a similar relation holds
approximately, $D_p\approx T_a\gamma_p$ with an ``active temperature''
$T_a$ defined as a mechanical quantity. Again, it is then enough to
compute the force autocorrelation, which we do in the limits of small
and large tracers compared to the persistence length of the active
particles.

\subsection{Passive bath}
\label{sec:FF-passive}

\begin{figure}
  \begin{center}
    \includegraphics[width=\textwidth]{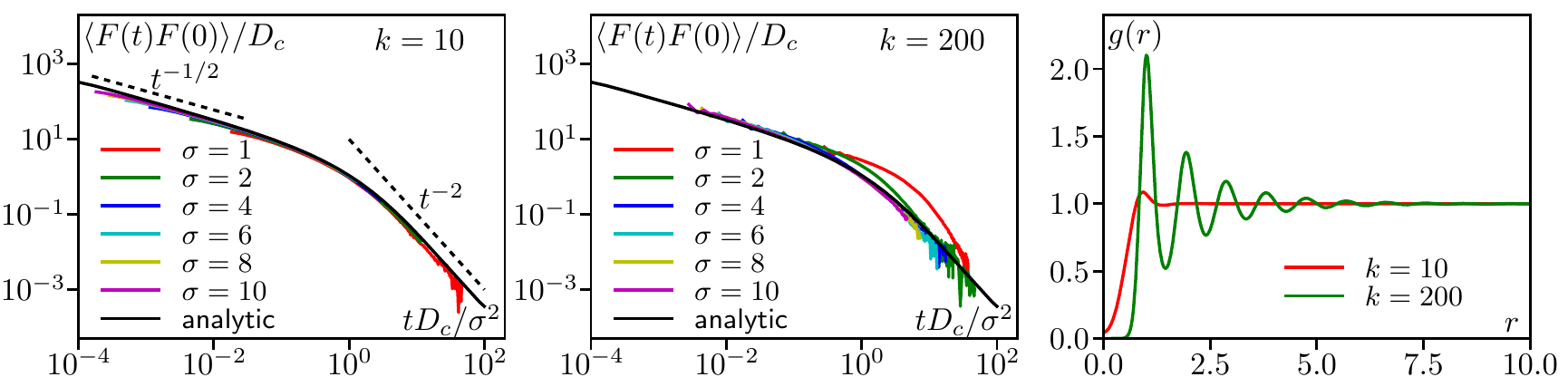}
    \caption{Force autocorrelation for a bath of PBP for varying
      $\sigma$ for soft interaction $k=10$ (left) and harder interaction
      $k=200$ (center). Simulations in 2d are compared to the analytic
      solution Eq.~(\ref{eq:FF-passive-int}). Right: Pair correlation
      in the fluid showing the increase of the correlation length when
      $k$ increases. $\rho_0=1$, $T=1$, time step $dt=0.003$ and
      system size $100\times 100$.}
    \label{fig:FF-passive-sigma}
  \end{center}
\end{figure}

Let us first compute the force autocorrelation
$\langle \vec F(t)\cdot \vec F(0)\rangle_B$ for non-interacting bath
particles. The element of force $\vec dF$ exerted on a surface element
$dS$ of the tracer around point $\vec r$ (with the origin at the
center of the tracer) is given by the ideal gas law
$\vec{dF}(0)=- T \rho_B(\vec r)dS \frac{\vec r}{|\vec r|}$ with
$\rho_B$ the density of bath particles around a fixed tracer. All
particles being independent, contributions to the force
autocorrelation come about only due to the same particle returning
multiple times to the surface. Denoting the transition probability for
a single particle in the presence of a tracer of size $\sigma$ as
$P_\sigma(\vec r',t | \vec r,0)$, the element of force exerted at time
$t$ on surface $dS'$ around $\vec r'$ is again given by the ideal gas
law
$\vec{dF}(t)=- T P_\sigma(\vec r',t | \vec r,0)dS' \frac{\vec r'}{|\vec
  r'|}$. Integrating over the surface $\mathcal S_T$ of the tracer
reads
\begin{equation}
  \label{eq:FF-passive1}
  \langle \vec F(t)\cdot \vec F(0)\rangle_B =\int_{\mathcal{S}_T} dS' \int_{\mathcal{S}_T} dS\  T^2 \frac{\vec r\cdot \vec r'}{|\vec r||\vec r'|} P_\sigma(\vec r',t | \vec r,0) \rho_B(\vec r).
\end{equation}
Using the spherical symmetry, we can reduce Eq.~(\ref{eq:FF-passive1})
to a particle starting on the $x$-axis at $\vec r=\sigma \vec e_x$ and
write, using that anywhere outside the tracer $\rho_B(\vec r)=\rho_0$,
the average density,
\begin{equation}
  \label{eq:FF-passive2}
  \langle \vec F(t)\cdot \vec F(0)\rangle_B =S_{d-1}\sigma^{d-1} \rho_0 T^2 \int_{\mathcal{S}_T} dS' \cos\theta P_\sigma(\vec r',t | \sigma \vec e_x,0).
\end{equation}
where $\theta$ is the angle between $\vec r$ and the $x$-axis and
$S_d$ is the solid angle of a $d$-dimensional sphere so that the area
of the tracer is $S_{d-1}\sigma^{d-1}$.  Finally, we rescale the
length by $\sigma$ in the integral, and recognize that
$P_\sigma(\vec r'/\sigma, t| \vec e_x,0)$ can be replaced by the solution of the
diffusion equation around a unit tracer with unit diffusion
coefficient, but with diffusively rescaled time
$(1/\sigma^d)P_1(\vec r, t D_B/\sigma|\vec e_x,0)$:
\begin{equation}
  \label{eq:FF-passive3}
  \langle \vec F(t)\cdot \vec F(0)\rangle_B = S_{d-1} \sigma^{d-2} \rho_0  T^2 \int_{\mathcal{S}_{d-1}} dS' \cos\theta P_1\left(\vec r',\frac{t D_B }{\sigma^2} | \vec e_x,0\right).
\end{equation}
The integral on the r.h.s. of Eq.~(\ref{eq:FF-passive3}) now runs over
the solid angle of a sphere $\mathcal{S}_{d-1}$ and is a dimensionless
function of the dimensionless time $\tilde t= t D_B/\sigma^2$ so that
we can write
\begin{equation}
  \label{eq:g}
 \fl\langle \vec F(t)\cdot \vec F(0)\rangle_B = S_{d-1} \sigma^{d-2} \rho_0  T^2 g(t D_B/\sigma^2); \qquad g(\tilde t)\equiv \int_{\mathcal{S}_{d-1}} dS' \cos\theta P_1(\vec r',\tilde t | \vec e_x,0)
\end{equation}
The function $g$ can be obtained by solving the diffusion equation
around the unit sphere. An explicit expression in terms of Bessel
functions is obtained for $d=2$ in \ref{sec:FF2d}.  From
Fig.~\ref{fig:FF-passive-sigma}, we see that is displays a power-law
behavior $g(\tilde t)\sim \tilde t^{-1/2}$ at short times crossing
over at $\tilde t\approx 1$ to $g(\tilde t)\sim \tilde t^{-2}$.

If the tracer is large enough, the previous calculation also applies
straightforwardly to interacting particles, once we recognize that the
only hydrodynamic mode in the bath is the density of
particles~\cite{Hess1983}. Indeed, on scales longer than the bath's
correlation length and time, the bath's particle density follows the
diffusion equation $\partial_t \rho_B=D_c\nabla^2 \rho_B$ with a
collective diffusion coefficient $D_c$.  On this scale, this is a full
description of the system, which is thus equivalent to non-interacting
particles with a diffusion coefficient $D_c$.  One can then repeat the
previous derivation, using that the mechanical pressure on the tracer
is now $\Pi =D_c\rho_0/\mu_B$ and obtain
\begin{equation}
  \label{eq:FF-passive-int}
  \langle \vec F(t)\cdot \vec F(0)\rangle_B=S_{d-1}\sigma^{d-2} \rho_0 T \frac{D_c}{\mu_B} g\left(\frac{D_c t}{\sigma^2}\right).
\end{equation}
Compared to Eq.~(\ref{eq:g}), note that only one of the prefactor $T$
was converted in $D_c/\mu_B$ in Eq.~(\ref{eq:FF-passive-int}). Indeed,
following the reasoning that lead to Eq.~(\ref{eq:g}), the initial
element of force $\vec d\vec F(0)$ is proportional to $D_c/\mu_B$ but
the one at time $t$, which corresponds to a single particle returning
to the tracer, exerts a pressure proportional to $T$.

Let us now compare the predictions of Eq.~(\ref{eq:FF-passive-int})
with results from numerical simulations in 2d. We first look at the
effect of the tracer size $\sigma$ in Fig.~\ref{fig:FF-passive-sigma}
by showing the measured force autocorrelation rescaled as in
Eq.~(\ref{eq:FF-passive-int}) and comparing with the analytic
expression obtained for a non-interacting bath. For soft interaction
$k=10$ (left panel), the agreement is excellent. For harder
interactions $k=200$ (center panel), one observes a deviation for
small tracers. This is hardly surprising since, as shown in
Fig.~\ref{fig:FF-passive-sigma} (right), pair correlations in the bath
extend over a larger distance as $k$ increases. For $k=200$, we see
that correlations extend over $\approx 3$ particle radii, consistent
with the deviations from scaling, which is expected only when the
tracer is larger than the correlation length of the bath.

\begin{figure}
  \begin{center}
    \includegraphics[width=\textwidth]{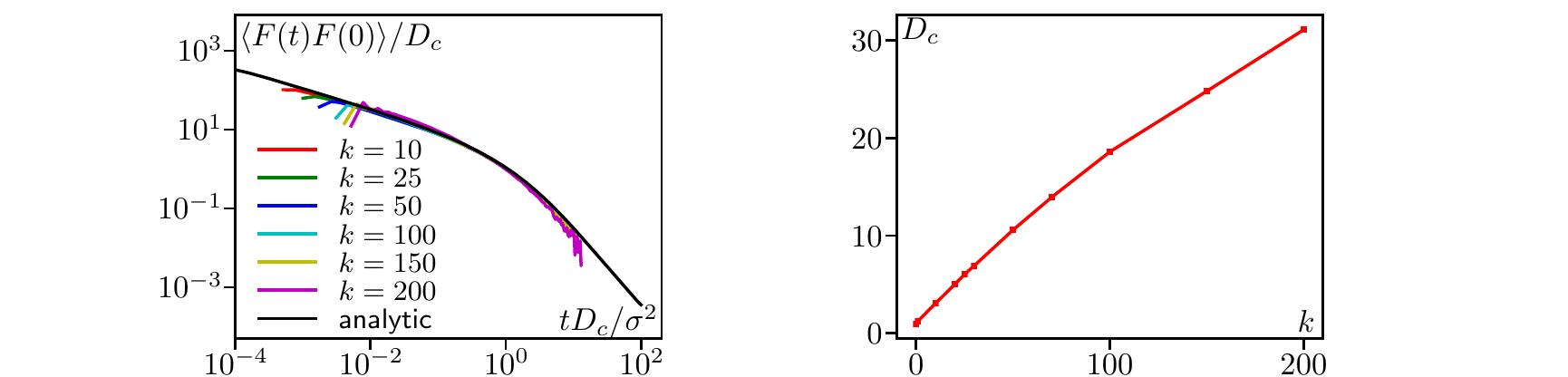}
    \caption{{\bf Left:} Force autocorrelation for a bath of PBP at
      fixed $\sigma=6$ for varying interaction strength $k$ showing a
      good collapse using the collective diffusion $D_c$. Parameters:
      $\rho_0=1$, $T=1$, $dt=0.003$ and system size $L=100$. {\bf
        Right:} $D_c(k)$ measured in relaxation experiments: We start
      from an inhomogeneous striped initial condition
      $\rho_B(x)=(1-\alpha)\rho_0$ if $x<L/2$ and
      $\rho_B(x)=(1+\alpha)\rho_0$ if $x>L/2$ with $\alpha=0.1$ and
      extract $D_c$ from the decay of the first Fourier mode
      $\vec q=(2\pi/L,0)$, such that
      $\hat\rho_B(\vec q,t)\propto e^{-q^2D_c t}$. Parameters: $L=100$,
      $dt=10^{-3}$, $\rho_0=1$.}
    \label{fig:FF-passive-Dc}
  \end{center}
\end{figure}

In Fig.~\ref{fig:FF-passive-Dc} we vary $k$ at a fixed tracer size
$\sigma=6$, larger than the correlation length for the range of $k$
values tested. We observe an excellent agreement in the whole
range. The collective diffusion coefficient $D_c$ used to rescale the
curves both in Fig.~\ref{fig:FF-passive-sigma} and
Fig.~\ref{fig:FF-passive-Dc} is computed independently in relaxation
experiments: Starting with an inhomogeneous initial condition (in our
case a stripe), the Fourier mode $\vec q$ of the density field decays
as $\hat\rho_B(\vec q,t)\propto e^{-D_c q^2t}$ which allows us to
extract $D_c$ from the rate of exponential decay. The resulting values
of $D_c$ are plotted in Fig.~\ref{fig:FF-passive-Dc} (right).  Note
that the collective diffusion coefficient increases by a factor of
$30$ when varying $k$ from $0$ to $200$ which thus provides a
significant test of Eq.~(\ref{eq:FF-passive-int}). For larger $k$
values, the correlation length diverges rapidly and our theory does
not apply anymore.

\subsection{Non-interacting active bath}
\label{sec:FF-active}

\begin{figure}
  \begin{center}
    \includegraphics[width=\textwidth]{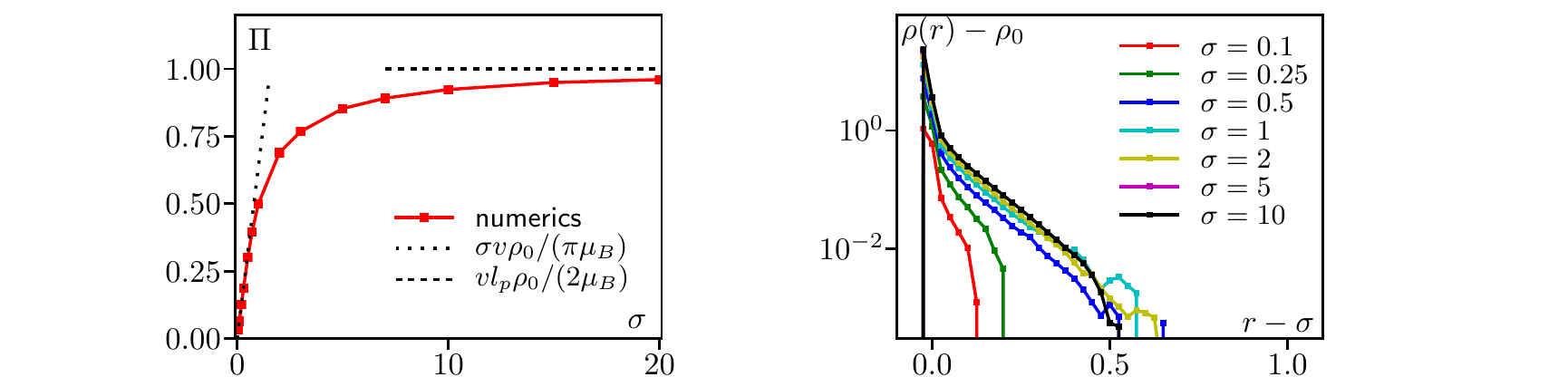}
    \caption{{\bf Left:} Pressure on the tracer due to a bath of
      non-interacting ABPs as a function of tracer size $\sigma$ from
      which the active temperature $T_a=\Pi/\rho_0$ is defined. The
      dashed and dotted line indicate the two asymptotic regimes
      considered in the text. Parameters: $v=2$, $l_r=1$,
      $\rho_0=1$. $dt=3\times 10^{-4}$ and $L=10$ for
      $\sigma<1$. $dt=3\times 10^{-3}$ and $L=60$ for $\sigma>1$. {\bf
        Right:} Density profile near the tracer for varying tracer
      size showing an exponential decay over a small
      distance. Parameters: $\rho_0=1$, $v=2$, $l_p=1$, $L=100$,
      $dt=3\times 10^{-3}$.}
    \label{fig:pressure-profiles}
  \end{center}
\end{figure}

Contrary to the passive case, one cannot resort to the Boltzmann
distribution to relate $\gamma_p$ and the force autocorrelation for
active particles. To proceed, let us consider non-interacting
particles for which the correlator in $\gamma_p$
can be written as an integral over the configuration of a single
particle (position $\vec r$ and orientation $\vec u$) instead of a
full microscopic configuration of the bath. This reads
\begin{equation}
  \label{eq:FlogNI}
\fl\langle \vec F(t)\cdot\nabla\log \pi_B(0)\rangle_B=\int d\vec r\int d\vec u\int d\vec r'\int d\vec u'\left[\vec F(\vec r',\vec u')\cdot \nabla \pi_B(\vec r,\vec u) P^A_\sigma(\vec r',\vec u',t | \vec r,\vec u,0)\right]
\end{equation}
where the spatial integrations run over all space and the orientation
ones over the unit sphere and $\vec F(\vec r',\vec u')$ is the force
exerted on the tracer by a particle at position $\vec r'$ with
orientation $\vec u'$. We have added a superscript $A$ to the
conditional probability $P^A_\sigma$ to distinguish it from the
passive case of Sec.~\ref{sec:FF-passive}. To make progress, we can
use that, for a hard tracer, $\nabla \pi_B$ is non-zero only close to
the surface where the probability distribution $\pi_B(\vec r,\vec u)$
goes rapidly from $0$ inside the tracer to its bulk value
$\rho_0/S_{d-1}$ (in the bulk all directions are
equiprobable). Although hard to prove rigorously, it is clear
numerically, as shown in Fig.~\ref{fig:pressure-profiles} where the
density $\rho_B=\int d\vec u\pi_B$ reaches $\rho_0$ exponentially on a
scale smaller than $\sigma$. On the contrary, $P^A_\sigma$ varies
smoothly over this range so that we can restrict the integration
$\int d\vec r$ to the surface of the tracer. The integration
$\int d\vec r'$ is also restricted to the surface since
$\vec F(\vec r',\vec u')$ vanishes elsewhere. Eq.~(\ref{eq:FlogNI})
thus simplifies to
\begin{equation}
  \label{eq:FlogNI2}
\fl\langle \vec F(t)\cdot\nabla\log \pi_B(0)\rangle_B\approx\int_{\mathcal{S}_T} dS \int d\vec u  \int_{\mathcal{S}_T} dS' \int d\vec u'\left[\vec F(\vec r',\vec u')\cdot \left( \frac{\rho_0}{S_{d-1}}\frac{\vec r}{|\vec r|}\right) P^A_\sigma(\vec r',\vec u',t | \vec r, \vec u,0)\right]
\end{equation}

We can now use the ideal gas law to relate Eq.~(\ref{eq:FlogNI2}) to
the force autocorrelation. The mechanical pressure due to
non-interacting active particles is a well-defined, measurable,
quantity~\cite{solon_pressure_2015}. Let us then introduce the
``active temperature'' $T_a$ such that the pressure on the tracer is
$\Pi=T_a\rho_0$. As in Sec.~\ref{sec:FF-passive}, the term $-T_a\rho_0\frac{\vec r}{|\vec r|}$ is then the element of force at time $t=0$ so that 
Eq.~(\ref{eq:FlogNI2}) reduces to the force autocorrelation up to a factor $1/T_a$, {\it i.e.}
\begin{equation}
  \label{eq:FlogNI3}
  \langle \vec F(t)\cdot\nabla\log \pi_B(0)\rangle_B\approx\frac{1}{T_a} \langle \vec F(t)\cdot \vec F(0)\rangle_B
\end{equation}
This implies that $D_p=T_a \gamma_p$ and both can be determined from
the force autocorrelation.

Note that the approximation sign in Eq.~(\ref{eq:FlogNI3}) comes from
the fact that $\nabla \pi_B(\vec r, \vec u)\neq 0$ over a finite range
near the surface of the tracer. For a passive fluid the dependence in
$\vec u$ disappears and $\nabla \pi_B(\vec r)=0$ strictly everywhere
except at the surface of the tracer so that Eq.~(\ref{eq:FlogNI3})
becomes an equality in the passive case. The temperature entering
the fluctuation-dissipation relation $D_p=T \gamma_p$ can thus be seen
as coming from the ideal gas law.

In addition to replacing $T$ by $T_a$, another important difference
with the passive case is that the conditional probability
$P^A_\sigma(\vec r',\vec u',t | \vec r, \vec u,0)$ appearing in the
correlators needs to be computed for an ABP instead of a PBP. In
general, this is not possible analytically and in the following we
look separately at the limits when the tracer is either very large or
very small compared to the persistence length of the ABPs.

\subsubsection{Large tracer}
On time scales larger than the persistence time $\tau_p=D_r^{-1}$, an
ABP looses its orientation and is thus effectively diffusing. For
large tracers $\sigma\gg l_p$, with $l_p=v\tau_p$ the persistence
length, the ABP is thus diffusive before it can leave ballistically
the tracer (which happens in a time $\sim \sigma/v$). In this case,
the motion approaches that of a passive particles and thus
$P^A_\sigma=P_\sigma$ (the dependence on $\vec u'$ and $\vec u$ in
$P^A_\sigma$ simply becomes irrelevant and can be integrated
out). Moreover, as shown in Fig.~\ref{fig:pressure-profiles}, the
pressure on the tracer approaches that on a flat
wall~\cite{solon_pressure_2015}, giving $T_a=\frac{v^2}{2D_r
  \mu_B}$. The force autocorrelation is then the same as in the
passive case Eq.~(\ref{eq:FF-passive3}) upon replacing $T$ by $T_a$ so that
\begin{equation}
  \label{eq:FF-active-large}
  \langle \vec F(t)\cdot \vec F(0)\rangle_B = S_{d-1} \sigma^{d-2} \rho_0  T_a^2 g\left(\frac{t D_B}{\sigma^2}\right).
\end{equation}
where $D_B\equiv \mu_B T_a$ is the bare diffusion coefficient of an
ABP.

Fig.~\ref{fig:FF-active} (left) verifies
Eq.~(\ref{eq:FF-active-large}) numerically in $d=2$. One observes that
indeed, upon increasing $\sigma$, the autocorrelation approaches the
same analytical solution as in Sec.~\ref{sec:FF-passive} for a passive
bath. The discrepancy at small $t$ is the signature of the finite
persistence of the active particles. However, as $\sigma$ increases,
the finite persistence becomes negligible in units of diffusive time
$\sigma^2/D_B$.

\subsubsection{Small tracer}
\begin{figure}
  \begin{center}
    \includegraphics[width=\textwidth]{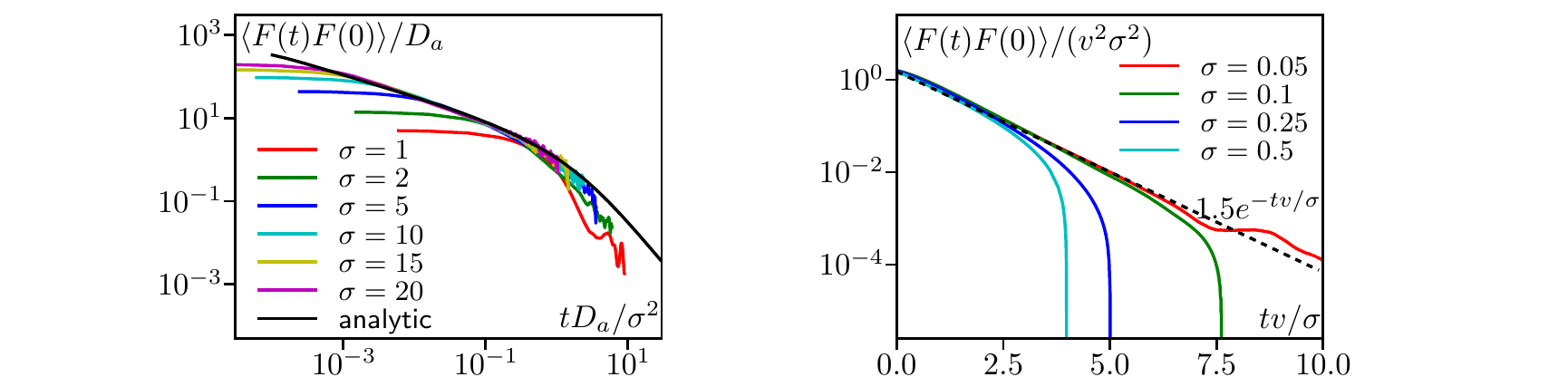}
    \caption{Force autocorrelation for a non-interacting active
      bath for varying $\sigma$. One observes to regimes with
      different scaling. For large $\sigma$ (left) the physics is the
      same as for a passive fluid. The physics at small $\sigma$
      (right), on the contrary, comes purely from the
      activity. Parameters: $v=2$, $l_r=1$. $L=150$ and
      $dt=3\times10^{-3}$ (left). $L=15$ and $dt=4\times10^{-4}$
      (right).}
    \label{fig:FF-active}
  \end{center}
\end{figure}

In the opposite limit, when $\sigma\ll l_p$, an ABP initially
in contact with the tracer will move far away before changing its
orientation. We then expect the relevant time scale to be the
ballistic time to leave the tracer $\sigma/v$. Moreover, in this
regime the active temperature $T_a$ depends on the tracer size
$T_a=\frac{v\sigma}{\pi \mu_B}$ in $d=2$ as shown numerically in
Fig.~\ref{fig:pressure-profiles} and
Ref.~\cite{smallenburg_swim_2015}. Consistent with this, the scaling
that is observed numerically in Fig.~\ref{fig:FF-active} (right) is of
the form
\begin{equation}
  \label{eq:FF-active-small}
  \langle \vec F(t)\cdot \vec F(0)\rangle_B =  \sigma^{d-2} T_a^2 \rho_0 h\left(\frac{t v}{\sigma}\right).
\end{equation}
and we find numerically in $d=2$ that $h(t)\approx 1.5 e^{-t}$. Note
that the scalings with $\sigma$, both in time and in amplitude
(because $T_a$ depends on $\sigma$) are different from the
large-tracer case above.

\section{Mobility and diffusivity of the tracer}
\label{sec:D-mu}

The coefficients $D_p$ and $\gamma_p$ appearing in the effective
dynamics of the tracer Eq.~(\ref{eq:tracer-dynamics-effective}) are
obtained by time-integrating the force autocorrelation. For the three
cases considered in Sec.~\ref{sec:FF}, integrating
Eq.~(\ref{eq:FF-passive-int}), Eq.~(\ref{eq:FF-active-large}) and
Eq.~(\ref{eq:FF-active-small}) yield respectively
\begin{eqnarray}
  \label{eq:gamma-passive}
  D_p= c \frac{\sigma^d T \rho_0}{\mu_B}=T\gamma_p \quad \, &{\rm (passive)}\\  
  \label{eq:gamma-active-large}
  D_p= c \frac{\sigma^d T_a \rho_0}{\mu_B}=T_a\gamma_p \quad \, &{\rm (active)}\, \sigma\gg l_p\\
  \label{eq:gamma-active-small}
  D_p= c' \frac{\sigma^{d} T_a \rho_0}{\mu_B}=T_a\gamma_p \quad \, &{\rm (active)}\, \sigma\ll l_p
\end{eqnarray}
with the constants $c= \frac{S_{d-1}}{d}\int_0^\infty g(t)dt$ and
$c'=\frac{\pi}{d}\int_0^\infty h(t)dt$. The three expressions above
are strikingly similar. The differences between these systems are
contained in the temperature $T$ or $T_a$ (which can depend on $\sigma$)
and the geometric constant $c$ or $c'$. In the passive case, the
collective diffusion coefficient $D_c$ that appears in the force
autocorrelation drops out at the integration. The dynamics of the
tracer is thus blind to the interactions in the bath.

\begin{figure}
  \begin{center}
    \includegraphics[width=\textwidth]{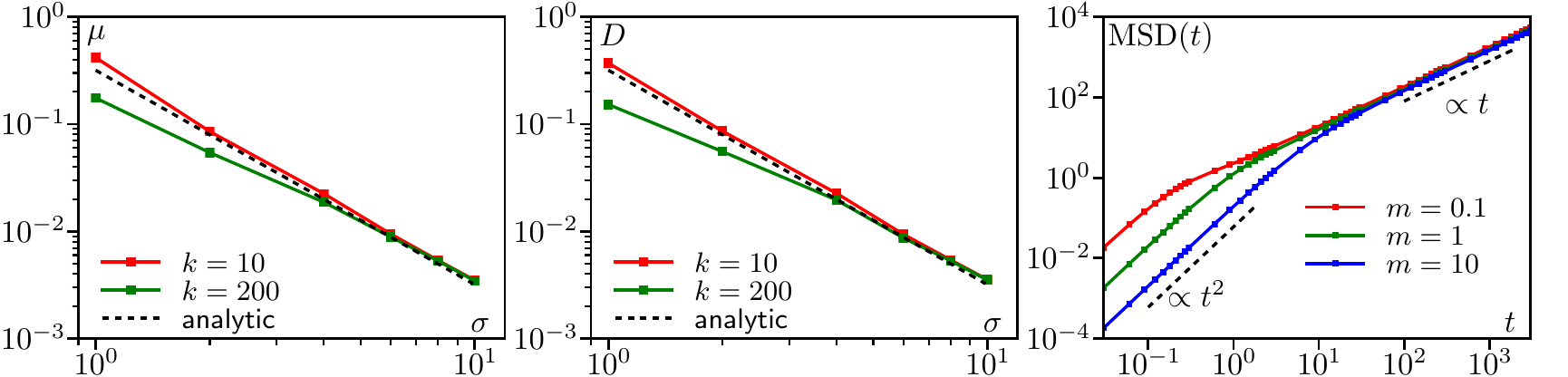}
    \caption{Interacting passive bath. {\bf Left:} Mobility of the
      tracer measured as the response to a pulling force small enough
      (here $f=0.03 \sigma^2$) to be in the linear regime. {\bf
        Center:} Diffusion coefficient measured by fitting the
      mean-squared displacement of the tracer at late times. {\bf
        Right:} Mean-squared displacement as a function of mass for
      three different values of the mass for $k=10$,
      $\sigma=1$. Parameters: $m=10$ (except on the right), $L=100$,
      $dt=3\times 10^{-3}$.}
    \label{fig:D-mu-passive}
  \end{center}
\end{figure}
From there, we can compute the diffusivity and mobility of the tracer
from Sec.~\ref{sec:eff-dynamics} which gives
\begin{equation}
  \label{eq:gamma-D-mu}
  \mu=\frac{1}{\gamma_T+\gamma_p};\qquad D=\frac{D_T+D_p}{(\gamma_T+\gamma_p)^2}.
\end{equation}

Let us first consider the case $\gamma_T=0$, meaning that the
surrounding fluid does not affect the tracer. This is probably not the
most common situation but provides a good test for the effect
of the bath on both $\mu$ and $D$. We show results with
$\gamma_T\neq 0$ at the end of this section.
For the passive bath or the active bath with $\sigma\gg l_p$, the
predictions from Eq.~(\ref{eq:gamma-passive}) and
Eq.~(\ref{eq:gamma-active-large}) are the same
\begin{equation}
  \label{eq:D-mu-passive-integrated}
  \mu=\frac{\mu_B}{c \rho_0\sigma^d}; \quad D=\frac{D_B}{c \rho_0\sigma^d}.
\end{equation}
In $d=2$ we obtain $c=\pi$ (see \ref{sec:FF2d}). The active case with
small tracers $\sigma\ll l_p$ gives
\begin{equation}
  \label{eq:D-mu-active-integrated}
  \mu=\frac{\mu_B}{c' \rho_0\sigma^d}; \quad D=\frac{\mu_B T_a}{c' \rho_0\sigma^d}.
\end{equation}
with $T_a=\sigma v/(\pi \mu_B)$ and $c'\approx 1.05$ in $d=2$.

We want to compare our predictions to numerical measurements of $D$
and $\mu$. The mobility is computed from its definition
Eq.~(\ref{eq:mu-def}) by applying a constant external force and
measuring the average velocity reached by the tracer. We ensured that
we are in the linear regimes by repeating the measurement for
different values of the external force. The diffusivity is fitted on
the late-time mean squared displacement of the tracer, in absence of
any external force.  In measuring $D$ and $\mu$, we make sure that we
have reached the limit where the tracer is heavy, the hypothesis used
in the calculation of Sec.~\ref{sec:eff-dynamics}. In practice, as
exemplified in Fig.~\ref{fig:D-mu-passive} (right) for the diffusion
coefficient, the variation with $m$ is no more than a few percent and
we found that using $m=10$ for all parameters is enough to ensure
convergence.

\begin{figure}
  \begin{center}
    \includegraphics[width=\textwidth]{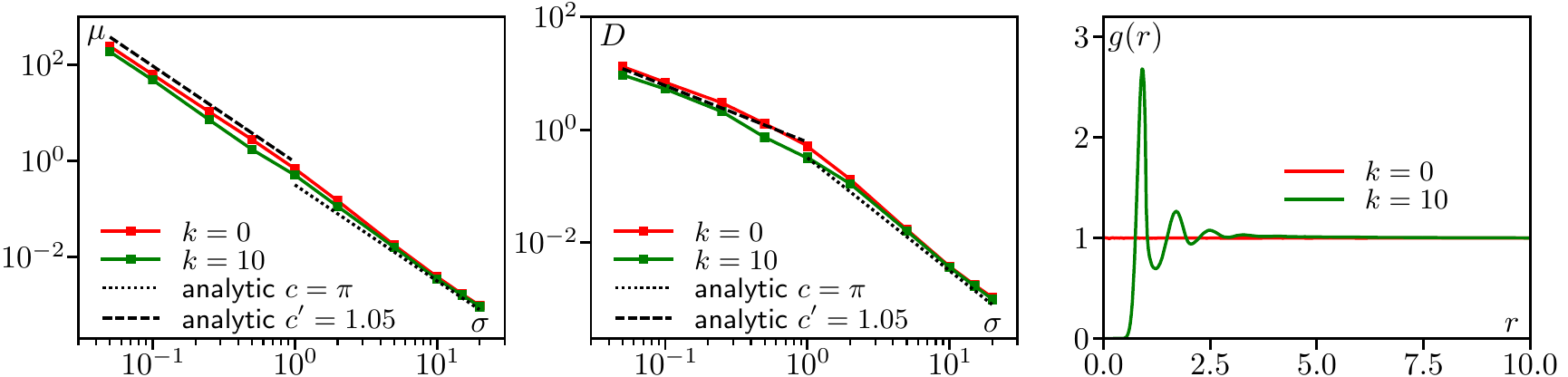}
    \caption{Non-interacting and interacting active baths. Mobility
      (left) and diffusion coefficient (center) measured as in
      Fig.~\ref{fig:D-mu-passive}. Parameters: $m=10$ and either
      $L=150$, $dt=3\times 10^{-3}$ for $\sigma<1$ or $L=15$,
      $dt=3\times 10^{-4}$ for $\sigma\ge 1$. {\bf Right:} Pair
      correlation in the bath showing that $k=10$ corresponds to a
      much higher interaction level than in the passive case (see
      text).}
    \label{fig:D-mu-active}
  \end{center}
\end{figure}

The predictions are compared without any fitting parameter to the
numerical simulations in Fig.~\ref{fig:D-mu-passive} for the passive
bath and Fig.~\ref{fig:D-mu-active} for the active one. For the
passive bath, the agreement is prefect within numerical accuracy
except for the expected deviation for tracers smaller than the
correlation length of the bath. In the active case of
Fig.~\ref{fig:D-mu-active} the agreement is again perfect when the
tracer becomes large. At small $\sigma$, we find a systematic
discrepancy of about $25 \%$ in the non-interacting case, which can be
explained by the approximation made in
Sec.~\ref{sec:FF-active}. Nonetheless, the theory captures the correct
order of magnitude and, for $D$, the correct change in behavior with
$D\propto\sigma^{-1}$ at small $\sigma$ and $D\propto\sigma^{-2}$ at
large $\sigma$.

If the active particles in the bath are interacting, we expect the
physics to be more complex since there are three length scales in the
system: the tracer size $\sigma$, the bath particle size and the
persistence length. However, in the limit of large $\sigma$, we expect
the same argument leading to Eq.~(\ref{eq:FF-passive-int}) in the
passive case to be applicable. Consistently, we observe
numerically in Fig.~\ref{fig:D-mu-active} that the values of $D$ and
$\mu$ do not depend on the interaction strength $k$ for large
$\sigma$. For smaller $\sigma$, their values depend slightly on $k$
but the asymptotic scalings $\mu\propto\sigma^{-2}$ and
$D\propto\sigma^{-1}$ are not modified.  Note that the value of the
interaction strength $k=10$ corresponds in the active case to a rather
strong interaction with several peaks in the pair correlation
function, as shown in Fig.~\ref{fig:D-mu-active} (right) while it was
close to the non-interacting limit in the passive case of
Fig.~\ref{fig:FF-passive-sigma}. This difference is not
surprising. Indeed, for the ballistic motion of ABPs the overlap
$\delta$ between two particles is of order $\delta\approx v/(\mu_B k)$
while for the diffusive PBPs $\delta\approx \sqrt{T/k}$ so that the
active particles become effectively stiffer more rapidly than passive
ones as $k$ increases.

\begin{figure}
  \begin{center}
    \includegraphics[width=\textwidth]{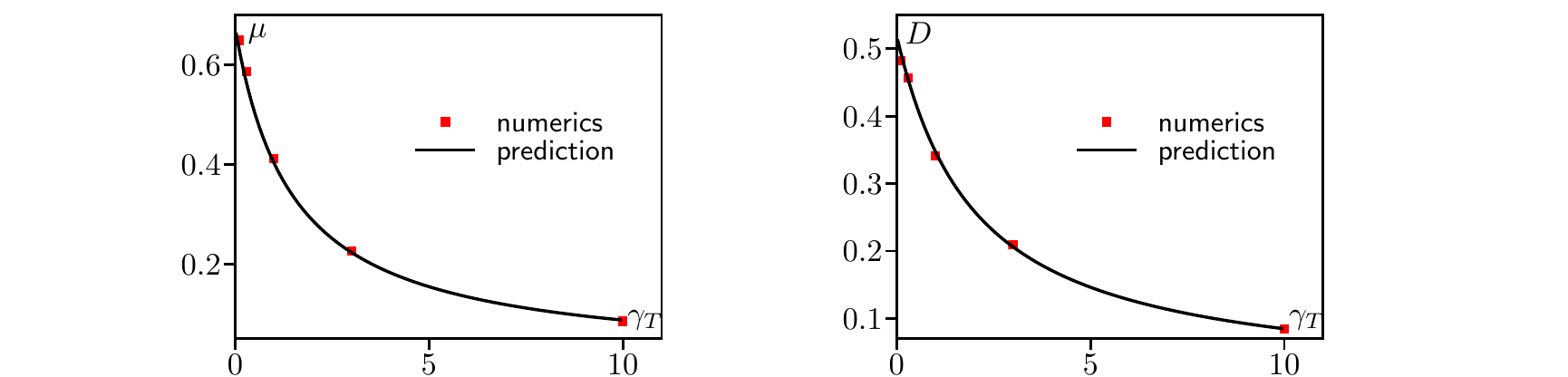}
    \caption{Effect of the external fluid friction with coefficient
      $\gamma_T$ on the mobility (left) and diffusivity (right) for an
      non-interacting active bath and a tracer size $\sigma=1$. On
      both panels, the prediction curve is that of
      Eq.~(\ref{eq:gamma-D-mu}), using the $\gamma_p$ and $D_p$
      measured at $\gamma_T=0$. Parameters: $k=0$, $m=10$, $L=150$,
      $dt=3\times 10^{-3}$.}
    \label{fig:gammaT}
  \end{center}
\end{figure}

Let us finally consider the case when the external fluid acts on the
tracer with a coefficient $\gamma_T\neq 0$. This is the most
relevant case experimentally since the most common active particles
such as bacteria or self-propelled colloids move in a fluid. The
mobility and diffusivity of the tracer is then given by
Eq.~(\ref{eq:gamma-D-mu}). To verify numerically this formula, we used
the values of $\gamma_p$ and $D_p$ measured when $\gamma_T=0$ and
extrapolate to $\gamma_T\neq 0$ using Eq.~(\ref{eq:gamma-D-mu}). We
see in Fig.~\ref{fig:gammaT} that the agreement with direct
measurements is perfect up to numerical accuracy.

\section{Discussion}
\label{sec:discussion}

In deriving the Einstein relation, our approach based on the effective
dynamics of the tracer makes clear that the temperature $T$ appearing
in the Einstein relation for a passive bath comes in through the ideal
gas law when computing the force autocorrelation in
Sec.~\ref{sec:FF}. Although an active bath is not characterized by a
single temperature, the mechanical pressure is a perfectly
well-defined quantity~\cite{solon_pressure_2015}. This leads us
naturally to define an ``active temperature'' $T_a$ as the quantity
appearing in the ideal gas law and therefore appearing also in the
fluctuation-dissipation relation $D_p= T_a \gamma_p$ albeit this is
only an approximation in the active case.

When the tracer is subject to both thermal noise and the active bath,
the temperatures associated to the two processes do not simply
add. Instead the coefficient in the $T_{\rm eff}$ such that
$D=\mu T_{\rm eff}$ reads from Eq.~(\ref{eq:betaeff})
\begin{equation}
  \label{eq:Teff}
  T_{\rm eff}=\frac{ T\gamma_T+T_a\gamma_p}{\gamma_T+\gamma_p}
\end{equation}
which features the damping coefficients due to the fluid and to the
bath, respectively $\gamma_T$ and $\gamma_p$. Different regimes are
then obtained depending on the parameters. When $\gamma_T\to 0$, as in
Fig.~\ref{fig:D-mu-active}, the fluid does not affect the tracer so
that $T_{\rm eff}=T_a$. When $\gamma_T\gg \gamma_p$ but
$T_a\gamma_p\gg T \gamma_T$, the damping is controlled by the fluid
but the diffusivity comes from the active bath. As discussed in the
introduction, this is the regime in which most experiments are
performed. One then has $T_{\rm eff}=T_a
\gamma_p/\gamma_T$. Ultimately, when $\gamma_T\gg \gamma_p$ and
$T\gamma_T\gg T_a \gamma_p$, we reach a passive limit with $T_{\rm eff}=T$.

Finally, let us remark that for large tracers the values of $D$ and
$\mu$ appear universal for both passive and active baths. In this
limit we obtained
\begin{equation}
  \label{eq:mu-D-large}
  \mu=\frac{\mu_B}{c \rho_0\sigma^d}; \quad D=\frac{D_B}{c \rho_0\sigma^d};  
\end{equation}
where the only effect of the bath is through the bare ({\it i.e.}
single particle) mobility and diffusion coefficient $\mu_B$ and $D_B$.
This is in stark contrast to simple equilibrium fluids, where the
Stokes-Einstein relation connects the diffusivity of a large, heavy
tracer to the bath's viscosity $D\propto 1/\eta$ (in 3D).  While there
is no viscosity for Brownian baths as momentum is not conserved and
therefore not relevant on hydrodynamic scales, one still might suspect
that collective properties of the bath affect a tracer's dynamics.
However, we have observed that large tracers are blind to the
collective transport properties of the bath, which can be accessed
only using tracers that are smaller than the correlation length of a
passive bath or the persistence length of an active one.  We believe
that our understanding of this issue would benefit from a
mode-coupling analysis like that carried out in
Ref.~\cite{Schofield1992} to derive the Stokes-Einstein relation from
the microscopic Hamiltonian dynamics, where it was shown that one must
actually include all higher-order modes to correctly extract the
Stokes-Einstein relation.

Note that in this article we have considered only pairwise interaction
between bath particles. For active particles, other types of
interaction such as quorum-sensing or nonreciprocal ones are possible
and more complex phenomena such as flocking and motility-induced phase
separation could happen in the bath. The dynamics  of a tracer in these
complex fluids is an open problem.

\ack

This work was partially supported by a grant from the Thomas Jefferson Fund, a program of FACE.

\appendix

\section{Derivation of the effective tracer dynamics using projection operators}
\label{sec:effective-bath}
In this Appendix, we review the projection operator technique for extracting the reduced equations for the tracer particle.  
The methodology is standard~\cite{Gardiner, Zwanzig,Espanol2002}, with our derivation following closely the original exposition for equilibrium fluids \cite{vankampen1986}.

The probability distribution for the joint position of the tracer  in its phase space $({\mathbf R},{\mathbf P})$ and the positions of the  overdamped Brownian bath particles $\{{\bf r}_i\}$ evolves according the Fokker-Planck equation~\cite{Gardiner}
\begin{equation}\label{eq:fp}
\partial_t P({\mathbf R},{\mathbf P},{\mathbf r}_i) ={\mathcal L}_T P({\mathbf R},{\mathbf P},{\mathbf r}_i) + {\mathcal L}_B  P({\mathbf R},{\mathbf P},{\mathbf r}_i), 
\end{equation}
where the tracer dynamics are generated by the Fokker-Planck operator
\begin{equation}
{\mathcal L}_T = -\frac{1}{m}\nabla_{\mathbf R}\cdot{\mathbf P}-\nabla_{\mathbf P}\left({\mathbf F}-\frac{\gamma_T}{m}\mathbf{P}\right)+D_T\nabla_{\mathbf P}^2
\end{equation}
and the bath dynamics are generated by one of two operators depending on whether they are passive  (PBP) or active (ABP)
\begin{equation}
{\mathcal L}_B = \left\{\begin{array}{ll}
-\mu_B\sum_i\nabla_{{\mathbf r}_i}\cdot\left({\mathbf f}_i-{\mathbf F}_i\right)+D_B\nabla_{{\mathbf r}_i}^2 & {\rm (PBP)} \\
-\mu_B\sum_i\nabla_{{\mathbf r}_i}\cdot\left({\mathbf f}_i-{\mathbf F}_i\right)-v_0\nabla_{{\mathbf r}_i}\cdot {\mathbf u}_i+D_r\partial_{\theta_i}^2 & {\rm (ABP)}
\end{array}\right. .
\end{equation}

Within the projector-operator formalism we assume that the bath relaxation time is fast compared to the tracer relaxation time. 
Thus, we expect that with the tracer fixed, the bath will quickly relax to its steady state distribution conditioned on the position of the tracer $\pi_B({\mathbf r}_i|{\mathbf R})$ given as the solution of
\begin{equation}
{\mathcal L}_B\pi_B({\mathbf r}_i|{\mathbf R})=0.
\end{equation}
We will assume that this distribution is unique, or, put another way, the null space of ${\mathcal L}_B$ is one dimensional.  We will also find it convenient to introduce a notation for steady-state averages with the tracer fixed as $\langle A\rangle_B=\int \prod_id{\mathbf r}_i A({\mathbf r}_i)\pi_B({\mathbf r}_i|{\mathbf R})$.

To exploit this separation of time-scales we introduce a projection operator that projects the bath distribution onto the fixed-tracer steady-state distribution
\begin{equation}
{\mathcal P}P({\mathbf R},{\mathbf P},{\mathbf r}_i)=\pi_B({\mathbf r}_i|{\mathbf R})\int \prod_id{\mathbf r}_iP({\mathbf R},{\mathbf P},{\mathbf r}_i)\equiv \pi_B({\mathbf r}_i|{\mathbf R}) \rho({\mathbf R},{\mathbf P}),
\end{equation}
as well as the orthogonal projector ${\mathcal Q}={\mathcal I}-{\mathcal P}$, where ${\mathcal I}$ is the identity operator. 
Note, that as is required for a useful projection operator, it projects away the bath dynamics
\begin{equation}\label{eq:Pprop}
{\mathcal L}_B{\mathcal P}={\mathcal P}{\mathcal L}_B=0,
\end{equation}
which can be verified from the definitions of the operators.

We next apply ${\mathcal P}$ and ${\mathcal Q}$ to (\ref{eq:fp}) from the left, and insert the identity operator ${\mathcal I}={\mathcal P}+{\mathcal Q}$ to the right of the Fokker-Planck operators to obtain the pair of equations for the relevant part ${\mathcal P}P$ and irrelevant part ${\mathcal Q}P$,
\begin{eqnarray}\label{eq:Pp}
&\partial_t{\mathcal P}P={\mathcal P}{\mathcal L}_T{\mathcal P}P+{\mathcal P}{\mathcal L}_T{\mathcal Q}P\\
&\partial_t{\mathcal Q}P={\mathcal Q}{\mathcal L}_B{\mathcal Q}P+{\mathcal Q}{\mathcal L}_T{\mathcal P}P+{\mathcal Q}{\mathcal L}_T{\mathcal Q}P,
\end{eqnarray}
where we have suppressed the arguments of the probability distribution to avoid cluttering the equations.
Typically at this point one solves the equation for the irrelevant part ${\mathcal Q}P$ formally and substitutes it back into the equation for the relevant part ${\mathcal P}P$~\cite{Zwanzig}.
To obtain a manageable equation, one then typically makes the uncontrolled Markov approximation.
Here, we will take a slightly more circuitous  route that gives the same results, but has the advantage of making the approximations more clear.

To this end, let us formally introduce a small parameter $\epsilon\ll1$ into the equation for the irrelevant part ${\mathcal Q}P$ to make explicit the time-scale separation between fast bath and slow tracer dynamics:
\begin{equation}\label{eq:Qeps}
\partial_t{\mathcal Q}P=\frac{1}{\epsilon}{\mathcal Q}{\mathcal L}_B{\mathcal Q}P+{\mathcal Q}{\mathcal L}_T{\mathcal P}P+{\mathcal Q}{\mathcal L}_T{\mathcal Q}P.
\end{equation}
At the end, we will set $\epsilon=1$.
We can now look for a perturbative solution of the form
\begin{equation}
{\mathcal Q}P = q^{(0)}+\epsilon q^{(1)}+\cdots.
\end{equation}
Substituting this into (\ref{eq:Qeps}), we can now solve order by order in $\epsilon$.
At lowest order we find
\begin{equation}
{\mathcal Q}{\mathcal L}_B q^{(0)}=(1-{\mathcal P}){\mathcal L}_B q^{(0)}={\mathcal L}_B q^{(0)}=0,
\end{equation}
upon using (\ref{eq:Pprop}).
This equation implies that $q^{(0)}$ is in the null space of ${\mathcal L}_B$.  However, by definition ${\mathcal Q}P$ is orthogonal to the null space of ${\mathcal L}_B$.  Thus, the only solution is $q^{(0)}=0$.
At the next order we have
\begin{equation}
{\mathcal Q}{\mathcal L}_Bq^{(1)}+{\mathcal Q}{\mathcal L}_T {\mathcal P}P=0.
\end{equation}
We can solve this formally, to get the first nontrivial solution for the irrelevant part
\begin{equation}\label{eq:Qapprox}
{\mathcal Q}P \approx q^{(1)} = {\mathcal Q}\int_0^\infty ds\ e^{{\mathcal L}_B s}{\mathcal Q}{\mathcal L}_T{\mathcal P}P,
\end{equation}
now setting $\epsilon=1$ as it is no longer needed.

Having approximated the irrelevant part of the dynamics, we can derive a closed equation for the relevant dynamics by substituting (\ref{eq:Qapprox}) into (\ref{eq:Pp}),
\begin{equation}
\partial_t{\mathcal P}P={\mathcal P}{\mathcal L}_T{\mathcal P}P+{\mathcal P}{\mathcal L}_T{\mathcal Q}\int_0^\infty ds\ e^{{\mathcal L}_Bs}{\mathcal Q}{\mathcal L}_T{\mathcal P}P.
\end{equation}
We now turn to evaluating each term using the definitions of the operators.
The first term represents the Eulerian part of the dynamics:
\begin{eqnarray}
\fl
{\mathcal P}{\mathcal L}_T{\mathcal P}P&={\mathcal P}\left[-\frac{1}{m}\nabla_{\mathbf R}{\mathbf P}-\nabla_{\mathbf P}\left({\mathbf F}-\frac{\gamma_T}{m}\mathbf{P}\right)+D_T\nabla_{\mathbf P}^2\right]\pi_{\rm B}\rho\\
&={\mathcal P}\pi_B\left[-\frac{{\mathbf P}}{m}(\nabla_{\mathbf R}\ln\pi_B)-\frac{1}{m}\nabla_{\mathbf R}{\mathbf P}-\nabla_{\mathbf P}\left({\mathbf F}-\frac{\gamma_T}{m}\mathbf{P}\right)+D_T\nabla_{\mathbf P}^2\right]\rho\\
&=\pi_B\left[-\frac{{\mathbf P}}{m}\langle \nabla_{\mathbf R}\ln\pi_B\rangle_B-\frac{1}{m}\nabla_{\mathbf R}{\mathbf P}-\nabla_{\mathbf P}\left(\langle {\mathbf F}\rangle_B-\frac{\gamma_T}{m}\mathbf{P}\right)+D_T\nabla_{\mathbf P}^2\right]\rho\\
&=\pi_B\left[-\frac{1}{m}\nabla_{\mathbf R}{\mathbf P}+\frac{\gamma_T}{m}\nabla_{\mathbf P}\mathbf{P}+D_T\nabla_{\mathbf P}^2\right]\rho,
\end{eqnarray}
where in the last line we used the assumed spherical symmetry of the tracer-bath force to set $\langle {\mathbf F}\rangle_B=0$ and the fact that $\langle \nabla_{\mathbf R}\ln\pi_B\rangle_B=0$, as $\pi_B$ is normalized.
For the second dissipative term we evaluate the effect of the operators one at a time:
\begin{eqnarray}
&\fl{\mathcal P}{\mathcal L}_T{\mathcal Q}e^{{\mathcal L}_B s}{\mathcal Q}{\mathcal L}_T{\mathcal P}P \\
&\fl={\mathcal P}{\mathcal L}_T{\mathcal Q} e^{{\mathcal L}_B s}{\mathcal Q}\pi_B\left[-\frac{{\mathbf P}}{m}\langle \nabla_{\mathbf R}\ln\pi_B\rangle_B-\frac{1}{m}\nabla_{\mathbf R}{\mathbf P}-\nabla_{\mathbf P}\left(\langle {\mathbf F}\rangle_B-\frac{\gamma_T}{m}\mathbf{P}\right)+D_T\nabla_{\mathbf P}^2\right]\rho\\
&\fl={\mathcal P}{\mathcal L}_T{\mathcal Q} e^{{\mathcal L}_B s}\pi_B\left[-\frac{{\mathbf P}}{m}(\nabla_{\mathbf R}\ln\pi_B)-\nabla_{\mathbf P}{\mathbf F}\right]\rho\\
&\fl={\mathcal P}{\mathcal L}_Te^{{\mathcal L}_B s}\pi_B\left[-\frac{{\mathbf P}}{m}(\nabla_{\mathbf R}\ln\pi_B)-\nabla_{\mathbf P}{\mathbf F}\right]\rho\\
&\fl={\mathcal P}\left[-\frac{1}{m}\nabla_{\mathbf R}{\mathbf P}-\nabla_{\mathbf P}\left({\mathbf F}-\frac{\gamma_T}{m}\mathbf{P}\right)+D_T\nabla_{\mathbf P}^2\right]e^{{\mathcal L}_B s}\pi_B\left[-\frac{{\mathbf P}}{m}(\nabla_{\mathbf R}\ln\pi_B)-\nabla_{\mathbf P}{\mathbf F}\right]\rho\\
&\fl=\pi_B\left[\nabla_{\mathbf P}\left\langle {\mathbf F}e^{{\mathcal L}_B s}\nabla_{\mathbf R}\ln\pi_B\right\rangle_B\frac{{\mathbf P}}{m}+\nabla_{\mathbf P}\left\langle {\mathbf F}e^{{\mathcal L}_B s}{\mathbf F}\right\rangle_B\nabla_{\mathbf P}\right]\rho
\end{eqnarray}
Thus,
\begin{eqnarray}
\fl{\mathcal P}{\mathcal L}_T{\mathcal Q}\int_0^\infty ds\ e^{{\mathcal L}_B s}{\mathcal Q}{\mathcal L}_T{\mathcal P}P=\pi_B\left[\nabla_{\mathbf P}\cdot\hat\gamma_p\cdot\frac{{\mathbf P}}{m}+\nabla_{\mathbf P}\cdot\hat{D}_p\cdot\nabla_{\mathbf P}\right]\rho,
\end{eqnarray}
having identified
\begin{eqnarray}
\hat\gamma_p= \int_0^\infty\left\langle {\mathbf F}e^{{\mathcal L}_B s}\nabla_{\mathbf R}\ln\pi_B\right\rangle_B ds=\int_0^\infty \left\langle {\mathbf F}(s)\nabla_{\mathbf R}\ln\pi_B(0)\right\rangle_Bds\\
 {\hat D}_p= \int_0^\infty\left\langle {\mathbf F}e^{{\mathcal L}_B s}{\mathbf F}\right\rangle_B ds =  \int_0^\infty\left\langle {\mathbf F}(s){\mathbf F}(0)\right\rangle_B ds,
\end{eqnarray}
which we recognize as operator representations of steady-state correlation functions.
These expressions can be simplified using spherical symmetry to conclude that they are both proportional to the identify $\hat\gamma_p=\gamma_p{\hat I}$ and ${\hat D}_p=D_p{\hat I}$ and by replacing $\nabla_{\vec R}\to -\nabla$ due to the translational invariance of the bath steady state, thereby arriving at (\ref{eq:coeff_proj-operator}).

Finally, putting everything together, we find a closed equation for the tracer dynamics
\begin{equation}
\partial_t\rho = \left[-\frac{1}{m}\nabla_{\mathbf R}{\mathbf P}+\frac{\gamma_T+\gamma_P}{m}\nabla_{\mathbf P}\cdot\mathbf{P}+(D_T+D_p)\nabla^2_{\mathbf P}\right]\rho,
\end{equation}
where the equivalent Langevin equation is used in (\ref{eq:tracer-dynamics-effective}).

\section{Exact force-force correlation for a passive particle in 2D}
\label{sec:FF2d}

In this Appendix, we solve the 2D diffusion equation around a disk in order to find an explicit expression for $g(t)$ (\ref{eq:g}), which captures the time-dependence of the correlation function of the force on the tracer due to a bath of independent passive Brownian particles.

Determining the force correlation function is equivalent to finding the Green's function for a Brownian particle diffusing in an annulus of inner radius $\sigma$ and outer radius $L$.
For large $L$, the shape of the region should be immaterial, allowing us to compare the results of this analytical calculation to the simulations.
Denoting the position of the particle as $(r,\theta)$ in polar coordinates, we  obtain the Green's function as the solution of the diffusion equation for the time-dependent probability distribution $P(r,\theta,t)$,
\begin{equation}\label{eq:2Ddiff}
\partial_tP(r,\theta,t)=\nabla^2P(r,\theta,t)=\frac{1}{r}\partial_r\left[r\partial_rP(r,\theta,t)\right]+\frac{1}{r^2}\partial_\theta^2P(r,\theta,t),
\end{equation}
with delta-function initial condition at the point $(r_0,\theta_0)$ and no flux boundary conditions,
\begin{eqnarray}
P(r,\theta,0)=\delta(r-r_0)\delta(\theta-\theta_0)/r_0\\
\partial_r P(\sigma,\theta,t)=\partial_r P(L,\theta,t)=0,\quad P(r,0,t)=P(r,2\pi,t).
\end{eqnarray}
The solution can be obtained using an eigendecomposition
\begin{equation}\label{eq:Green}
P(r,\theta,t)=\sum_\lambda e^{-\lambda t}f_\lambda(r,\theta)f_\lambda(r_0,\theta_0),
\end{equation}
where the eigenfunctions $f_\lambda$ satisfy
\begin{equation}\label{eq:f}
-\lambda f(r,\theta)=\frac{1}{r}\partial_r\left[r\partial_rf(r,\theta)\right]+\frac{1}{r^2}\partial_\theta^2f(r,\theta).
\end{equation}
While the solution is well known in general, imposing the no-flux boundary conditions requires some delicate analysis, so we review the solution here.

To proceed we look for separable solutions $f_\lambda(r,\theta)=R(r)\Theta(\theta)$.
Substituting this ansatz into (\ref{eq:f}), we find that the pair of functions $R$ and $\Theta$ must satisfy
\begin{eqnarray}\label{eq:theta}
\partial_\theta^2\Theta(\theta)+l^2\Theta(\theta)=0\\
\label{eq:R}
r^2\partial_r^2R(r)+r\partial_rR(r)+\left(\lambda r^2-l^2\right)R(r)^2=0,
\end{eqnarray}
where the constants $\lambda$ and $l$ are fixed by the boundary conditions.
To satisfy the periodic boundary condition, $\Theta(0)=\Theta(2\pi)$, we find that $l=n$ with $n\in{\mathbb N}$ and then (\ref{eq:theta}) has two possible solutions
\begin{equation}
\Theta(\theta)=\left\{\begin{array}{c}\cos(n\theta) \\ \sin(n\theta) \end{array}\right. ,
\end{equation}
when $n\neq 0$ and is simply $\Theta(\theta)=1$ when $n=0$.

With $l=n$ an integer, we recognize that (\ref{eq:R}) is Bessel's equation, whose solution can be written in terms of $n$-th order Bessel functions of the first $J_n$ and second $Y_n$ kind,
\begin{equation}
R(r) = A J_n(\sqrt{\lambda} r)+B Y_n(\sqrt{\lambda}r),
\end{equation}
where the constants $A$ and $B$ are fixed by imposing the no-flux boundary conditions ($\partial_rR(\sigma)=\partial_rR(L)=0$):
\begin{eqnarray}\label{eq:bc1}
&A J'_n(\sqrt{\lambda}\sigma)+B Y'_n(\sqrt{\lambda}\sigma)=0\\
\label{eq:bc2}
&A J'_n(\sqrt{\lambda} L)+B Y'_n(\sqrt{\lambda}L)=0.
\end{eqnarray}
To have nontrivial solutions for $A$ and $B$ the two equations must be linearly dependent, which will only be true when the determinant vanishes,
\begin{equation}
\fl
\qquad\left|\begin{array}{cc}J'_n(\sqrt{\lambda}\sigma) & Y'_n(\sqrt{\lambda} \sigma) \\ J'_n(\sqrt{\lambda} L) & Y'_n(\sqrt{\lambda}L)\end{array}\right|=
J'_n(\sqrt{\lambda}\sigma)Y'_n(\sqrt{\lambda}L)-J'_n(\sqrt{\lambda} L)Y'_n(\sqrt{\lambda}\sigma)=0.
\end{equation}
This is the required condition to fix the eigenvalues $\lambda$.
To exploit this, let us define the zeros $\alpha_{nm}$ of the determinant equation as
\begin{equation}\label{eq:eigen}
J'_n(\alpha_{nm}\sigma)Y'_n(\alpha_{nm}L)-J'_n(\alpha_{nm}L)Y'_n(\alpha_{nm}\sigma)=0.
\end{equation}
Then the eigenvalues are $\lambda=\alpha_{nm}^2$, and our solution becomes
\begin{equation}
R(r) = A J_n(\alpha_{nm} r)+B Y_n(\alpha_{nm} r).
\end{equation}
We can now fix one of the constants using either of the boundary conditions (\ref{eq:bc1})-(\ref{eq:bc2}), as they are now linearly dependent.
The radial component of the eigenfunction is then
\begin{equation}
R_{nm}(r) = J_n(\alpha_{nm} r) Y'_n(\alpha_{nm} L)-Y_n(\alpha_{nm} r)J'_n(\alpha_{nm} L),
\end{equation}
as long $n>0$ and $m>0$.
However, when $n=0$, then $\lambda=0$ is a possible solution.  In this case, $R_{00}(r)=1$.
 
Putting it all together and normalizing, we find for our eigenfunctions
\begin{eqnarray}
&f_{nm}(r,\theta)=\frac{1}{\pi N_{nm}}R_{nm}(r)\times \left\{\begin{array}{c}\cos(n\theta) \\ \sin(n\theta) \end{array}\right. \quad &n,m\ge 1\\
&f_{0m}(r,\theta)=\frac{1}{2\pi N_{0m}}R_{0m}(r),  & n=0, m\ge 1\\
&f_{00}(r,\theta)=\frac{1}{\pi(L^2-\sigma^2)}, & n=m=0 
\end{eqnarray}
where we have introduced the normalization $N_{nm}=\int_\sigma^LR_{nm}(r)^2 r dr $. 
Substituting into (\ref{eq:Green}), we arrive at our final expression for the Green's function 
\begin{eqnarray}
\fl P(r,\theta,t|r_0,\theta_0,0)&=\frac{1}{\pi(L^2-a^2)}+\frac{1}{2\pi}\sum_{m\ge 1}e^{-\alpha_{0m}^2 t}\frac{R_{0m}(r)R_{0m}(r_0)}{N_{0m}}\\
&\fl\qquad\qquad+\frac{1}{\pi}\sum_{n,m\ge 1}e^{-\alpha_{nm}^2 t}\frac{R_{nm}(r)R_{nm}(r_0)}{N_{nm}}\left(\cos(n\theta)\cos(n\theta_0)+\sin(n\theta)\sin(n\theta_0)\right).
\end{eqnarray}

With the Green's function in hand, all that is left is to obtain $g(t)$ is to evaluate the integral in (\ref{eq:g}) for a tracer of radius $\sigma=1$ in two dimensions,
\begin{equation}
g(t)=\int_0^{2\pi} \cos(\theta)P(1,\theta,t|1,0,0) d\theta.
\end{equation}
We observe that $\cos(\theta)$ is orthogonal to every eigenfunction except when $n=1$. 
As a result, 
\begin{equation}
g(t) = \sum_{m\ge 1}e^{-\alpha_{1m}^2 t}\frac{R_{1m}(1)^2}{N_{1m}},
\end{equation}
Data in figures were generated by numerically approximating this sum using the first 1000 nonzero eigenvalues for parameter values $\sigma=1$ and $L=25$.

\bibliography{refs.bib}

\end{document}